\documentclass[apj]{aastex}
\usepackage{amsmath,amssymb,bm,mathtools}
\DeclareRobustCommand{\uvec}[1]{{%
  \ifcsname uvec#1\endcsname
     \csname uvec#1\endcsname
   \else
    \bm{\hat{\mathbf{#1}}}%
   \fi
}}
\usepackage{pslatex}
\usepackage{graphicx}
\shorttitle{MOF of High Redshift Quasars}
\shortauthors{Chakraborty et al.}
\begin{document}
\title{Mean Occupation Function of High Redshift Quasars from the Planck Cluster Catalog} 

\author{Priyanka Chakraborty\altaffilmark{1,2}, Suchetana Chatterjee\altaffilmark{1}, Alankar Dutta\altaffilmark{1,3}, Adam D.\ Myers\altaffilmark{4} }
\altaffiltext{1}{Department of Physics, Presidency University, 86/1 College street , Kolkata-700073, India}
\altaffiltext{2}{Department of Physics and Astronomy, University of Kentucky, Lexington, KY 40506, USA}
\altaffiltext{3}{Department of Physics, Indian Institute of Science, Bangalore-560012, India}
\altaffiltext{4}{Department of Physics and Astronomy, University of Wyoming, Laramie, WY 82072, USA}

\begin{abstract}
We characterise the distribution of quasars within dark matter halos using a direct measurement technique for the first time at redshifts as high as $z \sim 1$. Using the Planck Sunyaev-Zeldovich (SZ) catalogue for galaxy groups and the Sloan Digital Sky Survey (SDSS) DR12 quasar dataset, we assign host clusters/groups to the quasars and make a measurement of the mean number of quasars within dark matter halos as a function of halo mass. We find that a simple power-law fit of $\log\left <N\right> = (2.11 \pm 0.01) \log (M) -(32.77 \pm 0.11)$ can be used to model the quasar fraction in dark matter halos. This suggests that the quasar fraction increases monotonically as a function of halo mass even to redshifts as high as $z\sim 1$. 
\end{abstract}

\section{Introduction}
Quasars are tracers of highly active galaxies that can be used to analyze structure in the Universe at extremes of distance and luminosity. Since they act as a promising probe for exploring the large-scale cosmic web at high redshifts, cosmological studies of quasars have become increasingly popular since quasars were first discovered in the 1960s \citep{Schmidt63,Greenstein63}. The very high luminosity of quasars is believed to arise from the enormous amount of gas accretion on to the supermassive black hole (SMBH) at the centre of quasar host galaxies \citep{l&b69,soltan82}. Since galaxies are located inside dark matter halos and SMBH form inside massive galaxies, it is expected that the central SMBHs that drive quasars should have gravitational connections with their host dark matter halos. In particular, quantitative measurements of quasar clustering have provided hints about the host dark matter halos of quasars, and hence the connection between quasars, galaxy evolution, and large scale structure \citep[e.g.,][]{croometal04, porcianietal04, croometal05, gillietal05, myersetal06, myersetal07a, myersetal07b, coiletal07, shenetal07, shenetal09, shenetal13, wakeetal08,  rossetal09, coiletal09, hickoxetal09, hickoxetal11, krumpeetal10, krumpeetal12, krumpeetal15, allevatoetal11, cappellutietal12, whiteetal12, mountrichasetal13, koutoulidisetal13,  eftekharzadehetal15, eftekharzadehetal17}.

\begin{figure*}[t]
\begin{center}
\begin{tabular}{c}
        \resizebox{8cm}{!}{\includegraphics[angle=0]{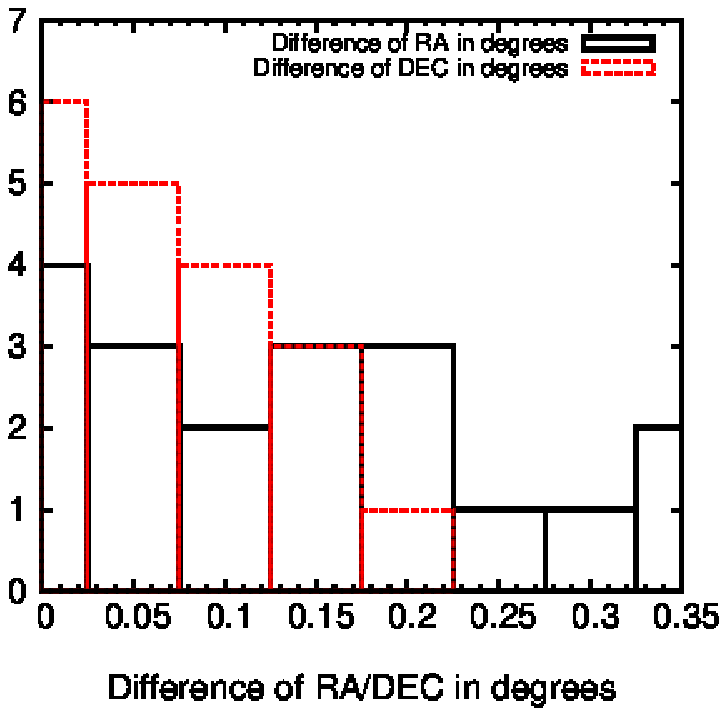}}\\
        \resizebox{9.5cm}{!}{\includegraphics[angle=270]{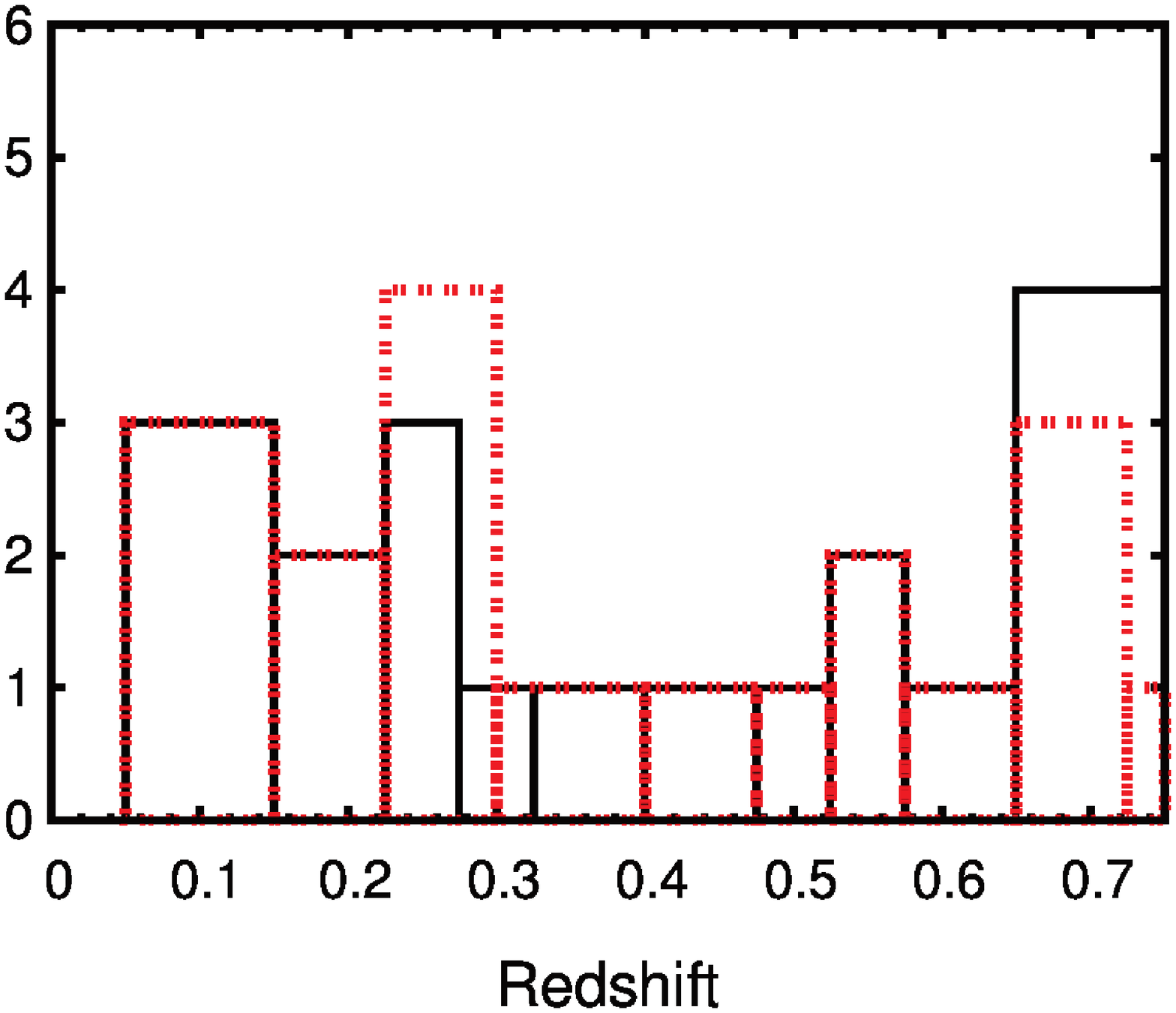}}
          \end{tabular}
 \caption{{\bf top panel:} The positions of the 19 clusters that match with a quasar, using a sample of 45,842 SDSS DR12 quasars and 913 PSZ1 clusters at redshifts $z \leq 1.0$. The histograms represent the differences in RA and DEC between the matched quasars and clusters. {\bf bottom panel:} The redshift distribution of matched quasars and clusters. Black solid and red dashed lines represent the redshift distributions of clusters and quasars respectively.}
\end{center}
\end{figure*}

At present, clustering measurements of quasars can be fully interpreted within the framework of the Halo Occupation Distribution \citep [HOD; e.g.,][] {wakeetal08,shenetal10,miyajietal11,starikovaetal11, allevatoetal11,richardsonetal12,richardsonetal13,k&o12,shenetal13}, which characterizes the bias of quasars in terms of individual host halo masses. The HOD is defined as the probability P(N$\mid$M) that a halo of mass M contains N objects of a given type (here, quasars), depending on the distribution of the objects within the halo \citep{z&w07}. Despite the promising application of HOD modeling to infer quasar clustering, choices of the correct HOD prescription remain largely degenerate when fitting the two-point correlation (2PCF) function to quasar clustering measurements \citep[e.g.][]{richardsonetal12, k&o12, chatterjeeetal13}. 

Given the failure of the 2PCF, alone, to distinguish between HOD models, additional observational constraints are critical to better constrain HOD parameters. Recent measurements have directly constrained the mean occupation function (MOF) of quasars at low redshift ($0.1\leq z \leq 0.3$) by cross-correlating quasars and clusters identified in Sloan Digital Sky Survey (SDSS) data \citep[C13 hereafter]{chatterjeeetal13}. This direct method is model-independent and is free from {\em a priori} theoretical biases. We note that in C13 we showed that at lower redshifts the mean occupation function of quasars strongly prefers a monotonically increasing function of halo mass. However in some HOD models it was suggested that the quasar occupation function falls off as a function of halo mass beyond a halo mass threshold of typically $10^{13-13.5}$M$_{\odot}$ \citep[e.g.,][]{k&o12}. Our measurements were in contradiction to those models but at a redshift where the quasar number densities are typically very low.

In this work, we perform a similar measurement to obtain the observational MOF of quasars extending to $z\leq 1.0$ to check if the C13 findings are consistent with quasar fraction measured at higher redshifts and if they are indeed monotonically increasing function of halo mass. To do this we directly match quasars drawn from the SDSS with the PlanckSZ catalog \citep{plancksz15} of galaxy clusters to obtain the fraction of quasars within dark matter halos. Estimating the direct observational MOF for quasars at a range of redshift intervals allows the MOF of quasars to be determined both directly and through estimates of the spatial correlation function, allowing the degeneracy in HOD models to be broken.

The paper is organized as follows. In \S 2 we describe our datasets and methodology. We present and discuss our results in \S 3. Throughout the paper, we have assumed a flat $\Lambda$CDM cosmology with $\Omega_{\rm m} = 0.28$, $\Omega_{\Lambda} = 0.72$, and $h = 0.71$ \citet{p15}.

\section {Datasets and Methodology}
We utilise the SDSS-DR12 quasar catalog \citep{sdss15} and the Planck Sunyaev-Zeldovich (SZ) \citep{plancksz15} cluster data for our analysis. In the next sections we give a brief description of our datasets and outline our methodology for constructing the MOF.

\begin{figure}[t]
\begin{center}
\begin{tabular}{c}
          \resizebox{8cm}{!}{\includegraphics[angle=270]{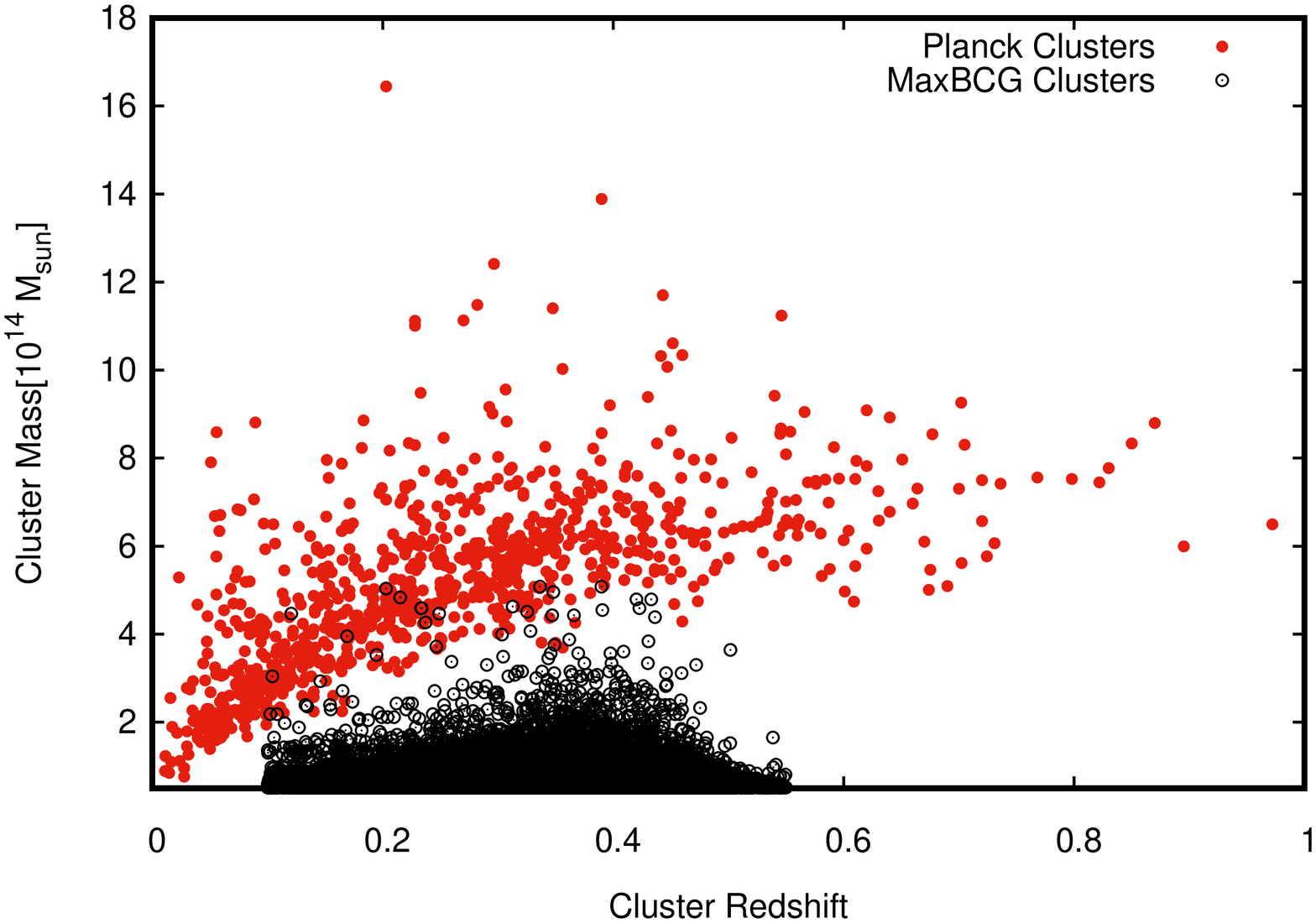}}\\
        \resizebox{8cm}{!}{\includegraphics{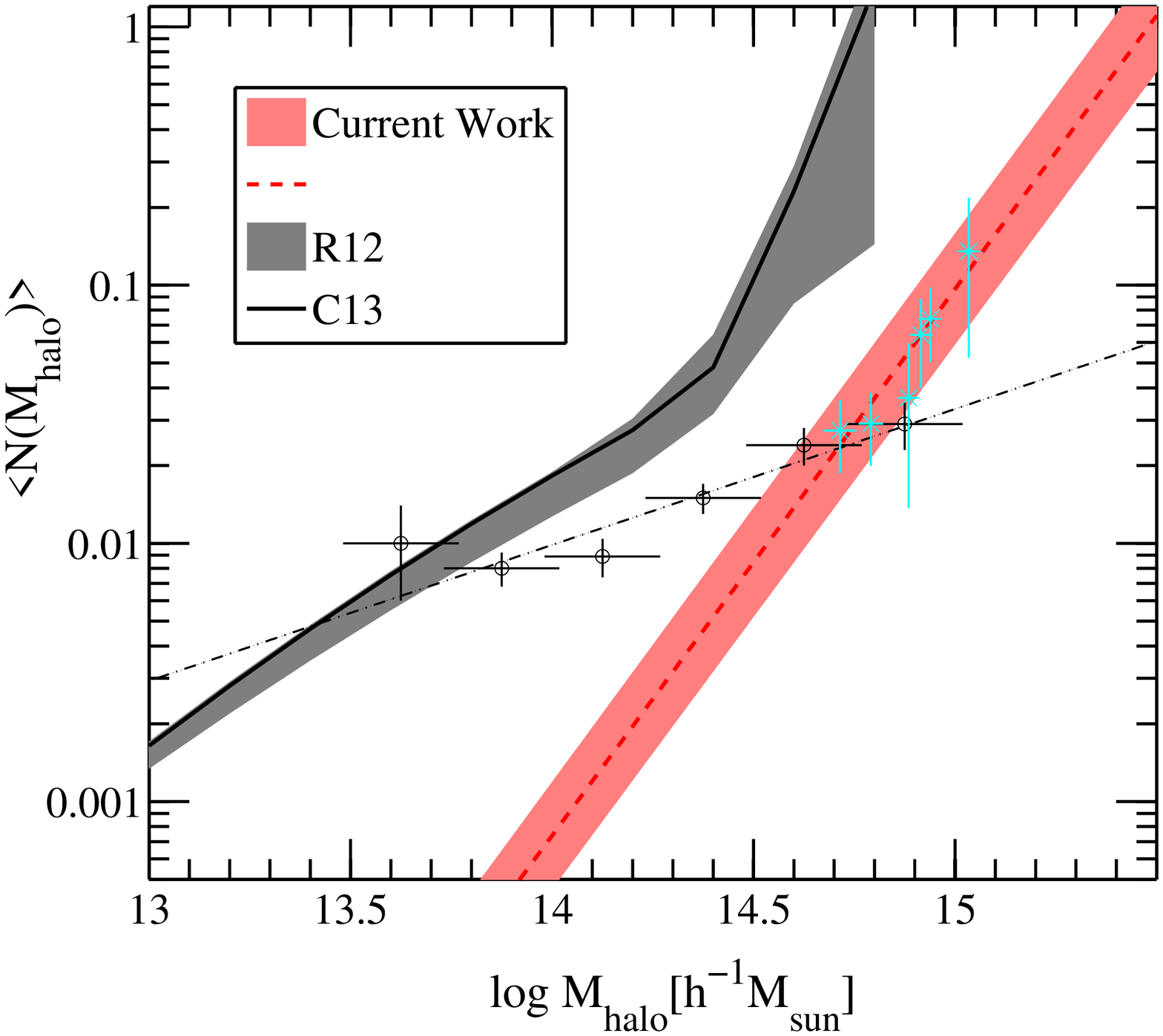}}
           \end{tabular}
 \caption{{\bf Top:} The redshift and mass distributions of the MaxBCG (black circles) and Planck SZ samples (red dots). {\bf Bottom:} Mean Occupation Function of quasars: The black circles depict the C13 results and the cyan stars correspond to our measurements. C13 showed that the MOF increases monotonically with mass within the redshift range $0.1\leq z \leq 0.3$. Our measurements demonstrate that the MOF for quasars continues to monotonically grow with halo mass to higher redshifts ($z\leq 1.0$). The red dashed and the black dot-dashed lines represent the best-fit power-law for the current work ($2.11 \pm 0.01$) and for C13 ($0.53 \pm 0.04$), respectively. The red shaded area represents the error on our measurements in the current work. The black thick solid line along with the shaded area depict the mean occupation function derived from the 2PCF measurements of R12.  }
\end{center}
\end{figure}

\subsection{Datasets}
To assign the quasars to their host dark matter halos, we use the updated all-sky Planck catalog (PSZ1) obtained from the first 15.5 months of Planck observations. The updated Planck catalog contains a total of 1227 cluster candidates including the redshifts of 913 systems spanning the redshift range $0.01 \leq z \leq 0.97$. Of these 913 systems, 733 have spectroscopically confirmed redshifts and 180 have redshifts that have been photometrically derived. The redshifts of the clusters are obtained from archival data and follow-up observations undertaken by the Planck collaboration. The Planck SZ cluster sample spans more than a decade in mass, ranging from $7.6\times 10^{13} M_{\odot} \leq M_{500}\leq 1.64 \times 10^{15} M_{\odot}$ within a scaled radius of $R_{500}$, as derived from the $Y-M$ mass calibration \citep{plancksz15}. 

The SDSS DR12 catalog \citep{parisetal12} comprises a total of 297{,}301 quasars covering more than one-third of the sky and spans a redshift range of  $0.04 \leq z \leq 6.44$ with a median redshift of $2.1$. The absolute $i$-band magnitude of these quasars is in the range $30.28 \leq M_{i} \leq 22.0$. For our analysis we have chosen 45{,}842 quasars at redshifts of $z \leq 1.0$ to match the redshift range of the PSZ1 cluster sample. The matched quasar and cluster samples are shown in Fig.\ 1. 

\subsection{Methodology}

We followed the method introduced by \citet{hoetal09} and C13 to characterize the host dark matter halos of quasars. We constructed a cylinder with a base radius $\theta_{200}$ and a height 2$\Delta z$ centered on each galaxy cluster center. Here, $\theta_{200}$ is the angular equivalent of $R_{200}$ in projected space and $R_{200}$ is the radius at which the mean density of a cluster is 200 times the mean matter density of the Universe \citep{NFW96}. We estimate the base radius of the cylinder using the relationship between $M_{200}$ and $R_{200}$ from the virial theorem. \citet{hoetal09} compute the cluster mass in terms of the critical density ($\rho_{\rm c}$) of the Universe. We, instead, adopt the mass definition prescribed by C13, who verified that replacing $\rho_{\rm c}$ by $\rho_{\rm mean}$ (the mean density of the Universe) produces similar results. 

In the Planck sample, the cluster masses are provided in terms of $M_{500}$. We converted $M_{500}$ to $M_{200}$ using the relation $M_{200}/M_{500}=1.5$, which we estimated from the mass profile of \citet{NFW96}. We note that the differences in mass and the sizes thereof (radii of clusters) introduced by the choice of mass definitions or their variations in redshifts, hardly affects the measurement of the MOF since the sample sizes are small and the statistical uncertainties are large. We would like to note that these systematic issues would become important if we have a larger matched sample (of higher statistical power) of quasars and clusters in future surveys. 

We assign a quasar to a cluster host if $\mid \theta_{\rm q}-\theta_{\rm c} \mid \leq \theta_{200}$ and $\mid z_{\rm q}-z_{\rm c}\mid \leq \Delta z$. Here, $\theta$ and  $z$ denote position and redshift, and the subscripts ``c'' and ``q'' denote the cluster and the quasar, respectively.  $\Delta z$ is the uncertainty in the measurement of the redshifts, which mostly arises from a combination of the relative velocity between the quasar and the cluster center, and the error in measuring the quasar and the cluster redshifts.

C13 used a mock catalog to test for biases arising from choice of $\Delta z$, cluster radius and other parameters in the reconstruction of the MOF (see C13 for more details). Following C13 we adopt $\Delta z=0.03$. Here we want to note a significant difference of our cluster sample with that of the MaxBCG sample used in C13. The MaxBCG cluster sample was a photometric sample and the error on cluster redshifts was $0.01$ \citep{koesteretal07}. This error dominated the redshift error budget since the redshift error on the spectroscopically selected quasar sample is insignificant. In the planck catalog a large number of the clusters have spectroscopic redshifts. However a substantial number of our clusters do have photometric redshifts too. In stead of trying to construct a complicated redshift selection function we have adopted a flat $\Delta z=0.03$ corresponding to the highest redshift errors in our PSZ1 clusters. We have also varied our choice of $\Delta z$ to observe its effect on the measured mean occupation function. We note that our results are mildly sensitive to the choice of $\Delta z$ and the changes are lower than the statistical error on our measurement of the occupation function. 

We detected a total of $19$ quasars within a sample of $913$ host halos in the redshift range $z \leq 1.0$. This corresponds to a quasar number density of $6.1 \times 10^{-10}\,h^{3}\,{\rm Mpc}^{-3}$ in galaxy clusters at the relevant redshifts. Since we had very few quasars detected in clusters, we simply performed a volume average to get an estimate of the number density. It is important to note that in our selection each quasar is assigned a unique host, but a given host (cluster) can have multiple quasars. After obtaining the host halos of the quasars, we use the $M_{200}$ value of each host cluster to calculate the quasar fraction as a function of host dark matter halo mass.

\section{Results and Discussion}
Fig.\ 2 shows our measured quasar MOF, together with recent results from the literature. We assume that the error on quasar number counts is Poisson, which follows from the fact that the number distribution of quasars in each halo mass bin follows a sub-Poisson distribution in the presence of both central and satellite components \citep[C12 hereafter] {degrafetal11b, chatterjeeetal12}. C13 showed that the quasar MOF increases monotonically with mass within the redshift range ($0.1\leq z \leq 0.3$). Our measurements show that this increase in the quasar fraction with mass is observed out to even higher redshifts ($z\leq 1.0$). 

The MOF measurement by C13 utilised a sample of galaxy groups drawn from the MaxBCG catalog \citep{koesteretal07}. The huge scatter between optical-richness and cluster mass in this sample \citep[$33\%$;][]{rykoffetal12} introduced a substantial uncertainty when estimating the mass of quasar hosts. There is, on the other hand, a tight correlation 
between SZ flux and cluster mass for the PSZ1 dataset, meaning that the error in mass estimation for our measurement is substantially less \citep[$\leq 10\%$;][]{plancksz15} than for C13. 

Given that our measurements are reasonably uncertain in both dimensions of the $\log (M)-\log\left <N\right>$ plane, we use a 
Markov Chain Monte Carlo (MCMC) approach to fit a power-law model to our observed quasar MOF. \citet{hoggetal10} and \citet{robothametal16} discuss rigorous approaches to modeling data that have multidimensional Gaussian uncertainties. 
We outline the fitting algorithm that we adopted in Appendix A, from which we obtain a best-fit power-law model of

\begin{equation} 
\log\left <N\right> = (2.11 \pm 0.01) \log (M) - (33.77 \pm 0.11) 
\label{eqn:fit}
\end{equation} 

Earlier, we discussed how the lack of a physically motivated theoretical model for the quasar MOF can lead to degeneracies when attempting to use the 2PCF of quasars to constrain the quasar HOD.
The main purpose of our work in this paper was to break this degeneracy by directly measuring the quasar MOF. R12 measured a monotonically increasing MOF using the parameterization of C12, which provided
a reasonable fit to the 2PCF of SDSS DR7 quasars and that of X-ray selected active galactic nuclei \citep[AGN;][]{richardsonetal13}, as well as that of infrared selected quasars \citep{mitra16}. 
\citet{k&o12}, however, used the same dataset to infer that the quasar fraction monotonically decreases with host halo mass. 

As mentioned above, \citet{k&o12} and R12 used the same 2PCF function of Sloan quasars and used two different HOD parameterization to fit the data. Both the groups obtained similar host halo mass scales, while there was a difference in satellite fraction as well as the high mass tail of the HOD. The median redshifts of these two measurements were $z \sim 1.4$ respectively. These two modeling with substantially different HODs led C13 to perform a direct measurement of the HOD using cluster and quasar samples. The best-fit power-law slope for the MOF obtained by C13 strongly favors ($\geq \sim 10 \sigma$) a monotonically increasing quasar occupation function with host halo mass. 

While the measured MOF of C13 essentially favored the HOD prescription of R12, it was not sufficient to make strong inferences about it. This was because the 2PCF and 
the HOD analysis of R12 and \citet{k&o12} were perfomed at much higher redshift ($z\sim 1.4$) than for C13, and it is certainly possible that the quasar HOD evolves with redshift. Therefore, for a complete interpretation of the functional form of the quasar occupation function, we extend our MOF measurement to higher redshifts ($z\leq 1$). We note that our results apply to both higher redshift 
clusters {\em and} higher mass clusters than those used in C13 (see left panel of Fig.\ 2). 

In the current high redshift work \citet{k&o12} model has been even more strongly ruled out ($\geq \sim 50 \sigma$) since we do see a quasar fraction that is very significantly increasing with halo mass. We emphasize that both C13 and the current work uses a completely model independent technique to measure the fraction of quasars in dark matter halos as a function of halo mass. We thus infer that the direct measurement strongly rules out quasar MOF models where quasar fraction falls-off at the high mass tail of the halo mass function. 

It is interesting to note that the power-law slope ($2.11 \pm 0.01$) we measure for the MOF of our high-redshift and high-mass sample is significantly steeper than the low-mass low-redshift slope ($0.53 \pm 0.04$) of C13.  According to C12, the quasar occupation fraction in dark matter halos increases beyond a mass scale where the quasar occupation is dominated by satellites. This broken-power-law behaviour is observed if we visually compare the current work with that of C13 (see Fig.\ 2). We note that neither the current measurement nor the measurement of C13 alone provided any statistically significant constraint on the C12 model. It is useful to jointly fit the C12 model combining the measurements of C13 and the current work to put constraints on the C12 parameters. Our joint fit did prefer the C12 model, but with high degeneracies between the C12 parameters. We also emphasize that such an analysis would require combining the selection functions of the MaxBCG and PlanckSZ catalogs in a meaningful manner. We thus propose to do this comparison with the MaxBCG and the high redshift RedMapper catalog \citep{rykoffetal14} which are both optically selected cluster samples and have more uniformity in their selection functions.

Our direct measurements of the quasar MOF may appear to be quite different to the HOD constraints from R12 plotted in Fig.\ 2. However, we note that the HOD inferred in R12 (black shaded region in Fig.\ 2) is based on the C12 parameterization which proposes a softened step function based occupation fraction for the central quasars and a power law occupation of the satellites. R12 obtained a satellite occupation slope of $0.62^{+1.9}_{-0.1}$. Due to 
the small sample size of our quasars-in-clusters sample we do not have the scope to model the central and satellite occupation separately. 

We also note that the halo mass scales used in this study are likely to have substantial satellite occupation. Hence we used a simple power-law model to fit our data, which is mainly a characterstic of a satellite population. We also note that in the 2PCF measurements majority of the quasars were residing in halos in the mass scale $\sim 10^{12-12.5}M_{\odot}$ (preferred halo mass of quasars) and the quasar occupation at the high mass tail was a natural extrapolation of the measurement that is mostly sensitive to lower mass halos. In the direct work however, our goal is find the quasar fraction at the high mass tail.  This induces a difference in the number density of quasars in our samples and hence careful consideration is required while comparing these two measurements. 

From Fig.\ 2 we see that the inferred MOFs from our work and R12 agree within the statistical limits and the value of the power-law slope for satellite occupation is consistent with our results. We however emphasize that there are a number of factors that complicate a direct comparison between our work and that of R12. In particular, the majority of the redshifts for the PlanckSZ clusters are obtained using archival data. Hence, the selection function of PlanckSZ clusters is potentially biased by a heterogeneous, redshift-dependent mass completeness \citep{plancksz15}. 

In the current work, as well as in C13, we tried to obtain the MOF of optically luminous quasars selected from the SDSS. However, since the connection between the central engine of a quasar and its host halo is 
believed to be a universal phenomenon, results at other wavelengths should be strongly related to our MOF constraints for SDSS quasars. For example, \citet{richardsonetal13} obtained the MOF of X-ray-selected 
AGN from the 2PCF measurements of \citet{allevatoetal11} and showed that the C12 model is capable of explaining the occupation functions of both optically luminous quasars and of X-ray selected AGN. 
As a further example, recent work by \citet{singhetal17} suggests that the MOF derived from the 2PCF of X-ray selected quasars using the C12 model 
by \citet{richardsonetal13} is compatible with occupation predictions from the X-ray luminosity function of \citet{airdetal15}, 
once the selection function of the X-ray telescope is taken into account.

Multiple other results suggest that a model similar to that of C12 can explain the MOF of X-ray-selected AGN. \citet{allevatoetal12} investigated the observational mean HOD for X-ray selected AGN 
using data from the XMM-Newton telescope. They modeled their MOF quite similarly to C12, except for an extra free parameter for 
the central quasar fraction. Their measurement was restricted to halo masses of $\log M_{200}\leq 14.5$ and redshifts of $z\leq 1$. They derived a best-fit power law slope of $0.06^{+0.36}_{-0.22}$ and a C12 model slope 
of $0.22^{+0.41}_{-0.29}$. The direct measurements of \citet{allevatoetal12} were consistent with the 2PCF measurements by \citet{richardsonetal13}. 
Influenced by this work, \citet{devoraetal16} performed an HOD analysis of X-ray AGNs modeled as supermassive black holes harbored by merging galaxies.
Their fitted slope for major mergers ($\alpha_{s} =0.20 \pm 0.18$) closely agrees with the C12 model slope derived by \citet{allevatoetal12}. It is notable that the slopes obtained for the X-ray AGN population are flatter than the ones obtained for optically bright quasars. This might hint at a mass-luminosity relation corresonding to AGN activity. 

The C12 parameterization referred here, is derived from a cosmological hydrodynamic simulation \citep{dimatteoetal08} which models growth and feedback of supermassive black holes in a cosmological hydrodynamic simulation. However, the spatial resolution of current cosmological volume simulation are limited to hundreds of parsec, which is well above length scales associated with the structure of supermassive black holes. Hence growth and feedback from black holes are restricted to subgrid models in these simulations. Also due the smaller box size the simulation did not have bright quasars and the most massive halos were group scale halos. 

For isolated galaxy simulations, the large scale environments of AGN and their cosmological context are absent. Hence most of these hydrodynamic simulations rely on observational constraints to ensure the fidelity of their parameters in the model. Despite these limitations, hydrodynamic simulations can still provide us information on the overall evolutionary history of AGN with their host galaxies and dark matter halos.  Recently \citet{khandaietal15} ran a bigger volume simulation which has higher mass halos and bright quasars. One of the key questions regarding quasar HOD lies in its redshift evolution. Till date no theoretical work has been done in characterizing the redshift evolution of the AGN/quasar HOD. With the simulation results of \citet{khandaietal15} we propose to carry out a theoretical study on the redshift evolution of quasars. The input from the current work will be crucial in this context.

Knowledge of the quasar HOD at high halo mass scales has become crucial in order to interpret many observations \citep [e.g., the SZ effect from quasar feedback][] {crichtonetal16, d&c17}. Traditionally, an understanding of how quasars relate to their host dark matter halos has been obtained from clustering measurements via the 2PCF. Using HOD prescriptions to robustly model the 2PCF still serves as the most promising method for understanding quasar co-evolution through future quasar/AGN surveys. 
HOD prescriptions, however, require a choice for the form of the MOF of quasars, and multiple different choices can still remain consistent with the overall shape of the quasar 2PCF. Direct measurement of the MOF is an extremely useful technique  in this context, and it provides a completely model-dependent avenue by which 
to obtain the quasar fraction in dark matter halos. The scope for conducting MOF measurements should increase substantially with upcoming and future multi-waveband surveys. Using complimentary methods to constrain the MOF and HOD of quasars 
should ultimately disentangle the relationship between quasars and their host dark matter halos.

\section*{Acknowledgments}  
We thank the referee for some very useful comments which helped in improving the quality of the draft. SC acknowledges support from the University Grants Commission through the start-up grant, Presidency University through the FRPDF grant and the Department of Science and Technology, Government of India through the SERB Early Career Research grant. SC is grateful to the Inter University Center for Astronomy and Astrophysics (IUCAA) for providing infra-structural and financial support along with local hospitality through the IUCAA-associateship program. ADM was partially supported by NSF grants 1515404 and 1616168, NASA grant NNX16AN48G and by the Director, Office of Science, Office of High Energy Physics of the U.S. Department of Energy under Contract No.\ DE-AC02-05CH1123.

\bibliography{PC_mybib}{}
\bibliographystyle{apj}

\appendix  

\section{The Fitting Algorithm}

We infer the slope and intercept of a power-law model for the quasar MOF (see, e.g., the form of Eqn.\ \ref{eqn:fit}) following
the methods outlined in \citet{hoggetal10} and \citet{robothametal16}. As we fit for both mass and number of objects in log space (i.e.\ as we fit in the $\log (M)-\log\left <N\right>$ plane)
 the error distribution on each of our parameters of interest is non-Gaussian. Further, in general, slopes and intercepts in power-law fits can be highly correlated.
We therefore adopt a Bayesian MCMC fitting approach, which can better handle non-Gaussian errors and highly correlated variables.

One effective way to incorporate non-Gaussian error in a Bayesian framework is to think of the uncertainties as a superposition of many Gaussian distributions with different weights. For the present case, we have assumed that the data is sampled from a probability distribution that corresponds to a superposition of two Gaussians of equal weight. Both of the Gaussians have a mean corresponding to the observed value at each point. 
One of the Gaussians has a standard deviation corresponding to the positive uncertainty, while the other has a standard deviation corresponding to the negative uncertainty. Under this set of assumptions, the correlation matrix is

\[S_{ij} = 
\begin{bmatrix}
    \sigma_{x_{ij}}^2 & \sigma_{xy_{ij}}\\
    \sigma_{xy_{ij}} & \sigma_{y_{ij}}^2\\
    \end{bmatrix}
\]

\noindent for each data point $i$, where we have adopted a superposition of $j=2$ Gaussians to characterize the error on each datum. 

Following \citet{hoggetal10} we define the probability of obtaining our dataset as:

$$P(x_i,y_i|S_{i\{j\}_{1}^{k}},x,y)=\sum_{j=1}^{k} \frac{a_{ij}}{2\pi\sqrt{\det(S_{ij})}}\exp\left(-\frac{1}{2}(X-X_{ij}^{'})^TS_{ij}^{-1}(X-X_{ij}^{'})\right)$$

\noindent where $a_{ij}$ is the assigned weight factor. Normalization requires that $\sum_{j=1}^{k}a_{ij}=\sum_{j=1}^{k}\Delta x_{ij}=\sum_{j=1}^{k}\Delta y_{ij}=1$, so, in our
case, $a_{i1}=a_{i2}=\frac{1}{2}$. Note that, broadly speaking,

\[X=
\begin{bmatrix}
	x\\
	y\\
\end{bmatrix},
\,\,\,
X_{i}=
\begin{bmatrix}
	x_{i}\\
	y_{i}\\
\end{bmatrix} {\rm and}
\,\,\,
X_{ij}^{'}=X_{i}+
\begin{bmatrix}
	\Delta x_{ij}\\
	\Delta y_{ij}\\
\end{bmatrix}  ,
\] 

\noindent but we have fixed $\Delta x_{ij}$ and $\Delta y_{ij}$ (which represent the offsets of the Gaussians from each datum) to be zero throughout our analysis. 

Defining

\[\uvec{v}=
	\frac{1}{\sqrt{1+m^2}}
	\begin{bmatrix}
	-m\\
	1\\
\end{bmatrix}=
\begin{bmatrix}
	-\sin(\theta)\\
	\cos(\theta)\\
\end{bmatrix}
\]
\[\Delta_{i}=\sum_{j=1}^{k}\uvec{v}^T X_{ij}^{'}-b\cos(\theta)
\]
\[
\Sigma_{i}^2=\sum_{j=1}^{k}\uvec{v}^T S_{ij} \uvec{v},
\]

\noindent the likelihood function can then be written as
$$\ln(\mathcal{L})=K-\sum_{i=1}^{N}\frac{\Delta_{i}^2}{\Sigma_{i}^2}$$

\noindent where $K$ is an arbitrary constant and $N$ is the total number of data points. 

We multiply $\mathcal{L}$ by a flat, uninformative prior in order to calculate a posterior probability. Finally, we MCMC sample the space of this posterior using the Python package {\tt emcee} \citep{foremanmackey13}.
Again following \citet{hoggetal10} we express the free parameters of our fit as $(\theta,b_{\bot})$ rather than $(m,b)$. Here $b_{\bot}=b\cos(\theta)$ is the 
perpendicular distance (from the origin) of the intercept that we are trying to estimate. This approach causes the linear fit to treat all slopes equally. In Fig.\ 3, we display our
inferred distributions for the slope and the intercept. 

For this work, we initialized 50 MCMC walkers and allowed them each to make $10^{5}$ steps. We discarded the first $4\times10^{4}$ steps for all of the walkers as a ``burn-in'' phase,
based on an auto-correlation study of the walker position series. We stopped the burn-in when
the auto-correlation decayed to zero, indicating that the walker positions had become independent, as is desired for a Markov chain.

\begin{figure}[t]
\begin{center}
\begin{tabular}{c}
        \resizebox{12cm}{!}{\includegraphics{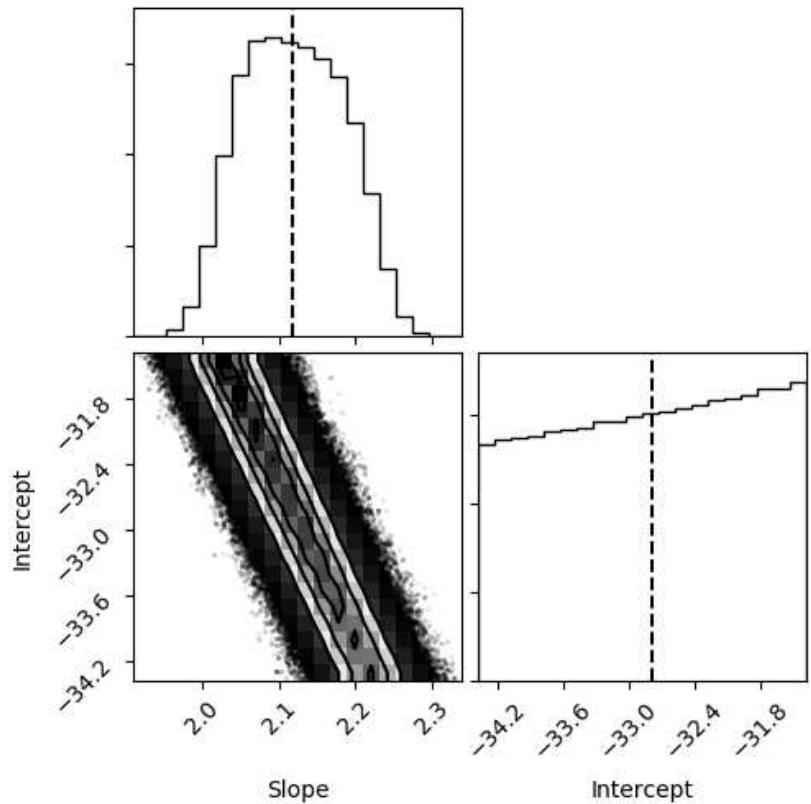}}
           \end{tabular}
 \caption{A sampling of the slope-intercept plane for our adopted power-law MOF. Plotted are the positions of 50 walkers each executing $10^5$ steps. The first $4\times10^{4}$ steps in each chain were excluded as a ``burn-in'' phase. 
The histograms depict the individual 1-D probability density distributions for the slope and the intercept walkers. The scatter plot in the lower-left corner is the joint distribution of the inferred slope and intercept, 
which demonstrates a strong correlation between the two parameters. }
\end{center}
\end{figure}

\end{document}